\documentclass{tMOP2e}

\usepackage{graphics}
\usepackage{graphicx}

\usepackage{ifthen} 
\usepackage{latexsym} 
\usepackage{substr} 

\newcommand{\ncd}{\newcommand}
\newcommand{\rcd}{\renewcommand}

\ncd{\ld}{\left} 
\ncd{\rd}{\right} 

\ncd{\el}{\ell}

\ncd{\ul}[1]{\underline{#1}} 
\ncd{\ut}[1]{\underline{\underline{#1}}} 
\ncd{\ct}{\ldots} 
\ncd{\tr}[1]{\mathrm{#1}} 
\ncd{\scs}{\scshape} 

\ncd{\oo}{\infty} 
\ncd{\ip}{\cdot} 
\ncd{\op}{\times} 
\ncd{\nd}{\frac} 
\ncd{\ml}{\times} 
\ncd{\mv}{\textrm{Max}} 
\ncd{\re}{\textrm{Re}} 
\ncd{\im}{\textrm{Im}} 
\ncd{\sr}{\sqrt} 
\ncd{\su}{\vert} 
\ncd{\gt}{\rightarrow} 
\ncd{\cv}{\ast} 
\rcd{\ss}{\ell} 

\ncd{\csch}{\mathrm{csch}} 
\ncd{\sech}{\mathrm{sech}} 

\ncd{\av}[1]{\ld| #1 \rd|} 
\ncd{\iv}[1]{\ld( #1 \rd)} 
\ncd{\pd}[2]{\partial_{#2} #1} 
\ncd{\ci}[2]{\ld[ #1, #2 \rd]} 
\ncd{\oi}[2]{\ld( #1, #2 \rd)} 

\ncd{\grad}{\nabla} 
\ncd{\divg}{\nabla \ip} 
\ncd{\curl}{\nabla \op} 

\ncd{\DD}[1]{\delta \iv{#1}} 
\ncd{\US}[1]{{\cal U} \iv{#1}} 

\ncd{\vt}[1]{\ul{#1}} 
\ncd{\dy}[1]{\ut{#1}} 
\ncd{\mr}[1]{\ld[ #1 \rd]} 

\ncd{\UV}[1]{\vt{u}_{#1}} 
\ncd{\NV}{\vt{0}} 
\ncd{\ND}{\dy{0}} 
\ncd{\ID}{\dy{I}} 
\ncd{\NM}{\mr{\dy{0}}} 
\ncd{\IM}{\mr{\dy{I}}} 

\ncd{\PV}{\vt{r}} 
\ncd{\PT}{\iv{\PV, t}} 
\ncd{\ZB}[1]{z_{#1}} 

\ncd{\Co}{c_{0}} 
\ncd{\Eo}{\epsilon_{0}} 
\ncd{\Mo}{\mu_{0}} 
\ncd{\No}{\eta_{0}} 

\ncd{\EF}{\vt{E}} 
\ncd{\BF}{\vt{B}} 
\ncd{\DF}{\vt{D}} 
\ncd{\HF}{\vt{H}} 
\ncd{\PF}{\vt{P}} 
\ncd{\SF}{\vt{S}} 
\ncd{\Sz}{S_{z}} 

\ncd{\FQ}{\omega} 
\ncd{\WL}{\lambda_{0}} 
\ncd{\AF}{\omega} 

\ncd{\EE}{\dy{\epsilon}_{r}} 
\ncd{\SD}[1]{\dy{\chi}_{#1}} 
\ncd{\RD}[1]{\dy{S}_{#1}} 
\ncd{\SDr}{\dy{\chi}_{ref}} 

\ncd{\AR}{\alpha} 
\ncd{\HP}[1]{\IfSubStringInString{c}{#1}{\Omega}{\IfSubStringInString{s}{#1}{\Gamma}{}}} 
\ncd{\SH}{h} 

\ncd{\LT}[1]{\chi_{#1}} 
\ncd{\OS}[1]{p_{#1}} 
\ncd{\NP}{p_{nl}} 
\ncd{\AP}[1]{N_{#1}} 
\ncd{\RW}[1]{\lambda_{#1}} 
\ncd{\RF}[1]{\omega_{#1}} 

\ncd{\Ec}{\hat{\epsilon}_{2}}
\ncd{\Ed}{\hat{\epsilon}_{d}}

\ncd{\Nc}{n_{2}}
\ncd{\Nd}{n_{d}}
\ncd{\Vpc}{v_{p2}}
\ncd{\Vpd}{v_{pd}}
\ncd{\Vgc}{v_{g2}}
\ncd{\Vgd}{v_{gd}}

\ncd{\FF}{\vt{F}} 
\ncd{\IF}{\vt{Q}} 
\ncd{\VD}{\dy{V}} 
\ncd{\WD}{\dy{W}} 

\ncd{\EV}{\mr{\EF}} 
\ncd{\FV}{\mr{\FF}} 
\ncd{\IV}{\mr{\IF}} 
\ncd{\SM}[1]{\mr{\SD{#1}}} 
\ncd{\RM}[1]{\mr{\RD{#1}}} 
\ncd{\VM}{\mr{\VD}} 
\ncd{\WM}{\mr{\WD}} 

\ncd{\dt}{\Delta t} 
\ncd{\dz}{\Delta z} 
\ncd{\bt}{\beta} 
\ncd{\NS}[1]{N_{#1}} 

\ncd{\LTc}[1]{\chi_{#1}^{c}} 
\ncd{\SMc}[1]{\mr{\SD{#1}^{c}}} 
\ncd{\IVc}{\mr{\IF^{c}}} 
\ncd{\WMc}{\mr{\WD^{c}}} 
\ncd{\WMi}{\mr{\WD'}} 

\ncd{\PE}{\psi} 
\ncd{\TC}{\tau_{0}} 
\ncd{\TD}{t_{d}} 
\ncd{\UE}{U} 
\ncd{\UT}{U_{t}} 

\ncd{\PW}{\vt{\varphi}} 
\ncd{\CF}{\omega_{car}} 
\ncd{\CP}{\phi} 
\ncd{\CW}{\lambda_{car}} 
\ncd{\TE}{\ld( s \rd)} 
\ncd{\TM}{\ld( p \rd)} 

\ncd{\FC}{f} 
\ncd{\ZR}{z_{r}} 
\ncd{\CEN}[1]{\zeta_{#1}} 
\ncd{\RMS}[1]{\sigma_{#1}} 
\ncd{\MOM}[2]{{\cal M}_{#1}^{\ld( #2 \rd)}} 
\ncd{\COR}{{\cal C}} 

\ncd{\TP}{\tau_{p}} 
\ncd{\TU}{\tau_{u}} 
\ncd{\TX}{\tau_{c}} 

\ncd{\SP}{c_{p}} 
\ncd{\SU}{c_{u}} 
\ncd{\SX}{c_{c}} 

\title{Quantification of optical pulsed--plane--wave--shaping by chiral sculptured thin films}

\author{Joseph B. Geddes III and Akhlesh Lakhtakia \\
CATMAS---Computational \& Theoretical Materials Sciences Group, \\
Department of Engineering Science \& Mechanics, \\
The Pennsylvania State University, \\
University Park, PA, 16802--6812, USA}

\date{12 April 2005}

\begin{document}

\maketitle

\begin{abstract}
The durations and average speeds of ultrashort optical pulses transmitted through chiral sculptured thin films (STFs) were calculated using a finite--difference time--domain algorithm. Chiral STFs are a class of nanoengineered materials whose microstructure comprises parallel helicoidal nanowires grown normal to a substrate. The nanowires are $\sim$10--300~nm in diameter and $\sim$1--10~$\mu$m in length. Durations of transmitted pulses tend to increase with decreasing (free--space) wavelength of the carrier plane wave, while average speeds tend to increase with increasing wavelength. An increase in nonlinearity, as manifested by an intensity--dependent refractive index in the frequency domain, tends to increase durations of transmitted pulses and decrease average speeds. The circular Bragg phenomenon exhibited by a chiral STFs manifests itself in the frequency domain as high reflectivity for normally incident carrier plane waves whose circular polarization state is matched to the structural handedness of the film and whose wavelength falls in a range known as the Bragg regime; films of the opposite structural handedness reflect such plane waves little. This effect tends to distort the shapes of transmitted pulses with respect to the incident pulses, and such shaping can cause sharp changes in some measures of average speed with respect to carrier wavelength. A local maximum in the variation of one measure of the pulse duration with respect to wavelength is noted and attributed to the circular Bragg phenomenon. Several of these effects are explained via frequency--domain arguments. The presented results serve as a foundation for future theoretical and experimental studies of optical pulse propagation through causal, nonlinear, nonhomogeneous, and anisotropic materials.
\end{abstract}



\section{Introduction} \label{S: Introduction}

An ultrashort optical pulse is a pulse of light of duration less than $\sim$10 optical periods, where the optical period is calculated from the dominant wavelength in the pulse spectrum. Such ultrashort pulses are generated with commonly implemented mode--locked lasers, and several schemes for compressing and amplifying them have been formulated~\cite{T.R.Gosnell-1991}. Optical pulses are used for micromachining~\cite{X.Liu-1997}, lidar~\cite{P.Rairoux-2000, M.Rodriguez-2004}, and quantum control and probing of wavepackets and chemical reactions~\cite{B.Kohler-1995, A.H.Zewail-1993}; and their use has been suggested for telecommunications~\cite{W.H.Knox-2000}. 

It is necessary to understand how ultrashort optical pulses propagate in materials in order to design new materials and devices to manipulate them. Earlier, we reported on the spatiotemporal evolution of ultrashort optical pulses as they propagate through complex nano--engineered materials called sculptured thin films (STFs). A STF is a unidirectionally nonhomogeneous, anisotropic medium whose morphology, which consists of an array of parallel shaped nanowires, is engineered by design~\cite{A.Lakhtakia-2005}. The morphology of STFs may break certain translational and rotational symmetries, and these films possess dispersive dielectric and possibly dispersive magnetic properties in the optical regime.  

\begin{figure}[t!]
\begin{center}
\scalebox{0.4}{\includegraphics{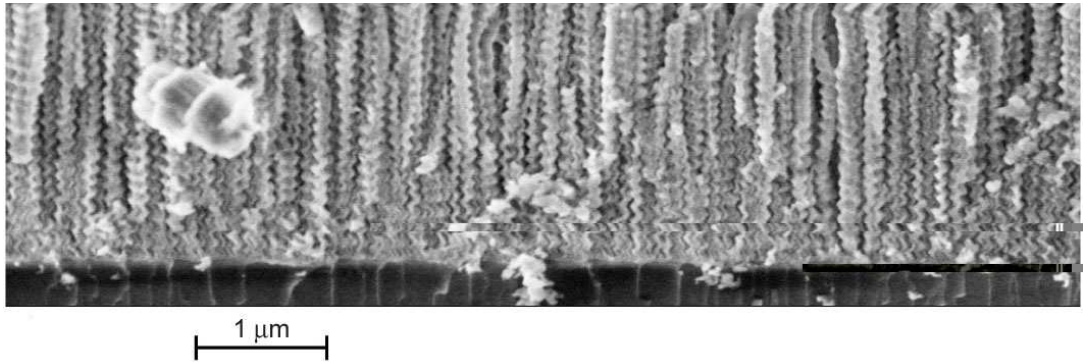}}
\end{center}
\caption{Scanning electron micrograph of a chiral sculptured thin film; note the periodic helical morphology (Courtesy: M. W. Horn, The Pennsylvania State University).} \label{F: Chiral STF}
\end{figure}

The circular Bragg phenomenon is exhibited by chiral STFs due to their periodic nonhomogeneity and structural chirality arising from the constituent nanowires being nanohelixes as shown in the electron micrograph of Figure~\ref{F: Chiral STF}. As circularly polarized monochromatic light propagates through vacuum, the electric field traces out a helix. If the handedness of this helix matches that of the nanohelixes in a chiral STF, and if the wavelength of the light lies within a range of wavelengths called the Bragg regime, the light normally incident on the chiral STF is mostly reflected; if the handedness of the helix is opposite, the reflection is little. In the time domain, the circular Bragg phenomenon is manifested as a pulse--bleeding phenomenon~\cite{J.B.GeddesIII-2000, J.B.GeddesIII-2001A}.

Previous research has shown that chiral STFs can affect the shapes, durations, and energy content of ultrashort optical pulses depending on the circular polarization state of the incident carrier plane wave~\cite{J.B.GeddesIII-2001A}. As the thickness of a chiral STF increases, the circular Bragg phenomenon ripens: the reflection first increases and then saturates, and differences in shape and duration between incident and transmitted pulses arise~\cite{J.B.GeddesIII-2001B}. Ultrawideband pulsed plane waves, known as videopulses, ultranarrowband rectangular pulsed plane waves, and two--dimensional pulsed beams can therefore also be shaped by chiral STFs~\cite{J.B.GeddesIII-2002, J.B.GeddesIII-2001C, J.B.GeddesIII-2005}. Pre-- and post--resonance Bragg regimes have also been elucidated~\cite{J.Wang-2002}.

So far, all research into the shaping of pulsed plane waves and pulsed beams by chiral STFs has been of a qualitative nature. The shapes of pulses reflected, refracted, transmitted, and diffracted by chiral STFs were predicted numerically, but measures of the extent of pulse shaping were not introduced. Our purpose here is to apply three measures of the pulse duration---the equivalent, root mean square, and correlation durations---and three measures of the average speed of the pulse---the peak speed, the center--of--energy speed, and the correlation speed---to the shaping of pulsed plane waves of different total energies, carrier wavelengths, and carrier polarization states after transmission through linear and nonlinear chiral STFs.

This work is important for four reasons. First, the measures of pulse shaping will allow direct comparisons between our theoretical conclusions and future experiments. Second, quantification of pulse shaping by complex materials like STFs will facilitate the design of STFs to shape pulsed plane waves and pulsed beams in controlled ways. Third, the algorithms we develop can be used to evaluate designs of STF--based devices; and fourth, they advance the state of the art of the time--domain analysis of electromagnetic pulse propagation to include propagation in causal, inhomogeneous, anisotropic, and nonlinear materials.

Parenthetically, by virtue of mathematical isomorphism we expect our results to hold, in general, for cholesteric liquid crystals (CLCs) too. We note that others have modeled optical propagation in noncausal nonlinear cholesteric liquid crystals (CLCs) in the frequency domain~\cite{P.Fu-1987} and with the slowly--varying envelope approximation~\cite{C.G.Avendano-2004}; propagation in linear CLCs has also been undertaken with the finite element method~\cite{Q.Hong-2003}. Second harmonic generation in nonlinear chiral STFs was also studied previously in the frequency domain~\cite{V.C.Venugopal-1998}. Our work distinguishes itself from these earlier studies on account of its causal time--domain treatment of the propagation of pulsed plane waves through nonlinear chiral STFs without the invocation of the frequency domain or slowly varying envelope approximation.

Our report on the quantification of the shaping of pulsed plane waves by chiral STFs proceeds as follows. The constitutive relations of a chiral STF, along with a finite--difference scheme for calculating the propagation of pulsed plane waves through vacuum and chiral STFs, are presented in Section~\ref{S: Theory}. Measures of the durations and average speeds of pulsed plane waves are also introduced in Section~\ref{S: Theory}. We discuss the main results of our work in Section~\ref{S: Results and discussion}, and present the main scientific and technological conclusions in Section~\ref{S: Concluding remarks}.

\section{Theory} \label{S: Theory}

This section describes the constitutive relations of spatially local, dielectric, structurally chiral STFs; our choice of ultrashort optical pulses; the numerical method we use to predict their propagation; and three measures each of pulse duration and speed.

\subsection{Maxwell curl postulates and constitutive relations}

We begin with the source--free time--domain Maxwell curl postulates
\begin{eqnarray}
\curl \EF \PT & = & - \pd{\BF \PT}{t} \, , \label{E: Faraday} \\
\curl \HF \PT & = & \pd{\DF \PT}{t} \, , \label{E: Ampere-Maxwell}
\end{eqnarray}
where $\EF$ and $\BF$ represent the primitive macroscopic electric and magnetic fields, respectively, while $\DF$ and $\HF$ are the corresponding induction fields. The electromagnetic field is a function of $\PV$, the position vector, and $t$, the time. We employ a cartesian coordinate system defined by the set of unit vectors $\ld\{ \UV{x}, \UV{y}, \UV{z} \rd\}$. The time derivative is denoted by $\pd{}{t}$.

A spatially local, dielectric, structurally chiral STF possesses the constitutive relations
\begin{eqnarray}
\DF \PT & = & \Eo \EF \PT + \PF \PT \, , \\
\BF \PT & = & \Mo \HF \PT \, ,
\end{eqnarray}
where $\Eo$ is the permittivity and $\Mo$ is the permeability of vacuum. The polarization $\PF$ is related to the electric field by the following equation:
\begin{equation} \label{E: Causal response}
\PF \PT  = \Eo \int_{0}^{t} \SD{e} \iv{\PV, t'} \ip \ld( 1 + \NP \av{\EF \iv{\PV, t - t'}}^{2}\rd) \EF \iv{\PV, t - t'} dt' \, .
\end{equation}
When $\NP = 0$, the film exhibits a linear response to an applied field. When $\NP \neq 0$, the response is third--order nonlinear; specifically, the material exhibits an intensity--dependent refractive index~\cite{V.C.Venugopal-1998, R.W.Hellwarth-1977}. The nonlinearity is thus isotropic. Whether linear or nonlinear, the chosen material does not respond instantaneously to the applied field. 

The dielectric susceptibility dyadic $\SD{e} \iv{\PV, t}$ contains the information about the nonhomogeneous, anisotropic, and temporally nonlocal properties of a chiral STF. Suppose the chiral STF occupies the region $\ci{\ZB{L}}{\ZB{R}}$. Then the dielectric susceptibility dyadic, which is null--valued in the region $z \notin \ci{\ZB{L}}{\ZB{R}}$, can be factored as:
\begin{equation}
\SD{e} \PT = \RD{} \iv{z - \ZB{L}} \ip \SDr \iv{t} \ip \RD{}^{-1} \iv{z - \ZB{L}} \, , \qquad z \in \ci{\ZB{L}}{\ZB{R}} \, , 
\end{equation}
where $\RD{}$ is a rotation dyadic that describes the morphology of the chiral STF along the $z$ axis (the axis of nonhomogeneity), and $\SDr \iv{t}$ is a reference susceptibility dyadic that describes, in accord with causality, the time--dependent and anisotropic response of the material. In this paper we treat unidirectionally nonhomogeneous STFs, and therefore exclude the recently fabricated bi-- and tri-- directionally nonhomogeneous films from consideration~\cite{M.W.Horn-2004A, M.W.Horn-2004B}.

The rotation dyadic takes the form
\begin{equation}
\RD{} \iv{z} = \RD{z} \iv{z} \ip \RD{y} \iv{\AR} \, ,
\end{equation}
where the dyadic
\begin{equation}
\RD{z} \iv{z}
= \cos \ld( \nd{\pi z}{\HP{c}} \rd) \ld( \UV{x} \UV{x} + \UV{y} \UV{y} \rd)
+ \SH \sin \ld( \nd{\pi z}{\HP{c}} \rd) \ld( \UV{y} \UV{x} - \UV{x} \UV{y} \rd)
+ \UV{z} \UV{z}
\end{equation}
depends on the structural half period $\HP{c}$ and the structural handedness parameter $\SH$, and the dyadic
\begin{equation}
\RD{y} \iv{\AR}
= \cos \AR \ld( \UV{x} \UV{x} + \UV{z} \UV{z} \rd)
+ \sin \AR \ld( \UV{z} \UV{x} - \UV{x} \UV{z} \rd)
+ \UV{y} \UV{y}
\end{equation}
depends on the angle of rise $\AR$. The chiral STF is either structually right--handed when $\SH = 1$ or structurally left--handed when $\SH = -1$. 

The reference susceptibility dyadic, which takes on an orthorhombic form due to the physical vapor deposition process by which chiral STFs are made~\cite{A.Lakhtakia-2005}, is defined as
\begin{equation}
\SDr \iv{t} 
= \LT{1} \iv{t} \UV{x} \UV{x} 
+ \LT{2} \iv{t} \UV{y} \UV{y} 
+ \LT{3} \iv{t} \UV{z} \UV{z} \, ,
\end{equation}
where
\begin{equation} \label{E: Lorentz model}
\LT{\ss} \iv{t} = \OS{\ss} \RF{\ss} \exp \iv{- \nd{\RF{\ss} t}{2 \pi \AP{\ss}}} \sin \iv{\RF{\ss} t} \US{t} \, , \quad \ss = 1, 2, 3,
\end{equation}
is determined both by the oscillator strengths $\OS{\ss}$ and by the parameters $\RW{\ss}$ and $\AP{\ss}$, which quantify the resonance wavelengths and absorption characteristics of the film; here, $\RF{\ss} = 2 \pi \Co / \RW{\ss}$ are the resonance angular frequencies and $\US{t}$ is the unit step function. The Lorentzian dependences~(\ref{E: Lorentz model}) are based on a classical material model that treats the charges in a material as undergoing damped oscillations when disturbed from equilibrium by an electric field~\cite{C.F.Bohren-1983}.

\subsection{Matrix partial differential equation}

We consider pulsed plane waves incident normally on chiral STFs, and hence there is no variation in the electromagnetic field along the $x$ and $y$ directions---i.e., $\pd{}{x} \equiv \pd{}{y} \equiv 0$. Upon writing the components of the electromagnetic field in a 6--vector as $\mr{\FF \iv{z, t}} = \ld[ E_{x}, E_{y}, E_{z}, H_{x}, H_{y}, H_{z} \rd]^{T}$, where the superscript $^{T}$ indicates the transpose, and substituting the constitutive relations into the Maxwell curl postulates (\ref{E: Faraday}) and (\ref{E: Ampere-Maxwell}), we obtain the matrix partial differential equation
\begin{equation} \label{E: MPDE}
\pd{\mr{\FF \iv{z, t}}}{t} = \Co \VM \pd{\mr{\FF \iv{z, t}}}{z} - (1 / \Eo) \, \pd{\mr{\IF \iv{z, t}}}{t} \, .
\end{equation}
In this equation,
\begin{equation}
\VM = \ld[ 
\begin{array}{cccccc}
0 & 0 & 0 & 0 & -\No & 0 \\
0 & 0 & 0 & \No & 0 & 0 \\
0 & 0 & 0 & 0 & 0 & 0 \\
0 & 1/\No & 0 & 0 & 0 & 0 \\
-1/\No & 0 & 0 & 0 & 0 & 0 \\
0 & 0 & 0 & 0 & 0 & 0
\end{array} \rd]
\end{equation}
is the vacuum propagation matrix, $\No = \sr{\Mo / \Eo}$ being the impedance of vacuum and $\Co = 1 / \sr{\Eo \Mo}$ the speed of light in vacuum, and the column vector 
\begin{eqnarray}
\mr{\IF \iv{z, t}} 
& = & \ld[ P_{x} \iv{z, t}, P_{y} \iv{z, t}, P_{z} \iv{z, t}, 0, 0, 0 \rd]^{T} \\
& = & \Eo \int_{0}^{t} \ld( 1 + \NP \av{\EF \iv{z, t - t'}}^{2} \rd) \mr{\WD \iv{z, t'}} \mr{\FF \iv{z, t - t'}} dt' \, .
\end{eqnarray}
In the previous equation,
\begin{equation}
\mr{\WD \iv{z, t}} = \ld[ 
\begin{array}{cc}
\mr{\SD{e} \iv{z, t}} & \NM \\
\NM & \NM
\end{array} \rd]
\end{equation}
contains the linear material properties, with $\mr{\ND}$ as the $3 \ml 3$ null matrix. We denote the matrix equivalents of vectors and dyadics by enclosing the corresponding symbols in square brackets.

Equation (\ref{E: MPDE}) can only be solved approximately, analytical solutions being virtually impossible for the materials being considered here. It was discretized with respect to both $z$ and $t$, using the leapfrog method. The details are available elsewhere~\cite{J.B.GeddesIII-2006}.

\subsection{Incident pulsed plane waves}

We chose the incident signal as a carrier plane wave that is amplitude--modulated by a pulse envelope $\PE \iv{t}$, i.e., at $z = 0$ we prescribe the boundary condition
\begin{equation} \label{E: PPW}
\mr{\FF \iv{0, t}} 
= \PE \iv{t} \ld[ \begin{array}{c}
\cos \iv{\CF t} \\
\cos \iv{\CF t + \CP} \\
0 \\
- \ld( 1 / \No \rd) \cos \iv{\CF t + \CP} \\
\ld( 1 / \No \rd) \cos \iv{\CF t} \\
0
\end{array} \rd] \US{t} \, ,
\end{equation}
where $\CF = 2 \pi \Co / \CW$ is the carrier angular frequency and $\CW$ is the carrier wavelength, while $\CP$ is a carrier phase. When $\CP = \pi / 2$, the carrier plane wave is right circularly polarized (RCP); when $\CP = - \pi / 2$, the carrier plane wave is left circularly polarized (LCP). The initial condition was chosen to be $\mr{\FF \iv{z, 0}} = \mr{\NV}$.

Even though we treat pulsed plane waves in this paper, it should be noted that no electromagnetic signal---or STF---can have an infinite transverse extent; indeed, all physical electromagnetic signals are pulsed beams. However, the pulsed--plane--wave approximation to a pulsed beam is appropriate when the beamwidth is much larger than the carrier wavelength (but still smaller than the transverse extent of the STF). The characteristics of two--dimensional pulsed beam propagation in chiral STFs have been reported elsewhere~\cite{J.B.GeddesIII-2005}.

We chose the pulse envelope $\PE \iv{t}$ to be gaussian, so that
\begin{equation} \label{E: Pulse envelope}
\PE \iv{t} = \sr{\nd{\No \UT}{\TC \sr{\pi}}} \, \exp \iv{- \nd{1}{2} \ld( \nd{t - \TD}{\TC} \rd)^{2}} \, , \quad t > 0 \, ,
\end{equation}
but $\PE \iv{t} = 0$ for $t \leq 0$. The time constant $\TC$ governs the duration of the pulse, while $\TD$ is the time delay which permits the whole pulsed plane wave to enter the computation domain. The prefactor on the right side of~(\ref{E: Pulse envelope}) is chosen so that 
\begin{equation}
\int_{-\oo}^{\oo} \Sz \iv{z, t} dt = \UT \, , 
\end{equation}
$\Sz = \UV{z} \ip \ld( \EF \op \HF \rd)$ being the axial component of the instantaneous Poynting vector. Hence, $\UT$ is the total energy per unit area of the incident pulsed plane wave. 

\subsection{Quantification of pulse durations and average speeds} \label{s: Quantification of pulse shapes and velocities}

We now turn to quantitative measures of pulse duration and average speed. Our goal is to define measures that are useful for experimental research or theoretical insight or both. As the chosen pulses are highly broadband and suffer appreciable distortion, the usual concept of group velocity~\cite{K.E.Oughstun-1994} is not meaningful for analysis except in frequency--domain arguments. More sophisticated measures were therefore devised.

The pulse duration is an important parameter because it determines the time scale for measurements or manipulations made with the pulse; pulses of shorter duration permit the measurement of events of shorter duration~\cite{R.Trebino-2000}. In order to define measures of pulse duration, we considered the $m^{\tr{th}}$ moment $\MOM{\FC, \xi}{m}$ of a scalar function $\FC \iv{\xi}$ with respect to variable $\xi$ on the interval $\ci{\xi_{a}}{\xi_{b}}$: 
\begin{equation}
\MOM{\FC, \xi}{m} \iv{\FC \iv{\xi}, \ci{\xi_{a}}{\xi_{b}}} = \int_{\xi_{a}}^{\xi_{b}} \xi^{m} \FC \iv{\xi} d\xi \, .
\end{equation}
The root mean square (RMS) deviation $\RMS{\FC, \xi}$ from the centroid
\begin{equation}
\CEN{\FC, \xi} \iv{\FC \iv{\xi}, \ci{\xi_{a}}{\xi_{b}}} = \nd{\MOM{\FC, \xi}{1} \iv{\FC \iv{\xi}, \ci{\xi_{a}}{\xi_{b}}}}{\MOM{\FC, \xi}{0} \iv{\FC \iv{\xi}, \ci{\xi_{a}}{\xi_{b}}}}
\end{equation}
of $\FC \iv{\xi}$ with respect to $\xi$ on the interval $\ci{\xi_{a}}{\xi_{b}}$ is defined as follows~\cite{R.N.Bracewell-2000}:
\begin{eqnarray}
\RMS{\FC, \xi} \iv{\FC \iv{\xi}, \ci{\xi_{a}}{\xi_{b}}} 
& = & \sr{\nd{1}{\MOM{\FC, \xi}{0} \iv{\FC \iv{\xi}, \ci{\xi_{a}}{\xi_{b}}}} \int_{\xi_{a}}^{\xi_{b}} \ld( \xi - \CEN{\FC, \xi} \rd)^{2} \FC \iv{\xi} d\xi} \, , \\
& = & \sr{\nd{\MOM{\FC, \xi}{2} \iv{\FC \iv{\xi}, \ci{\xi_{a}}{\xi_{b}}}}{\MOM{\FC, \xi}{0}\iv{\FC \iv{\xi}, \ci{\xi_{a}}{\xi_{b}}}} - \CEN{\FC, \xi}^{2} \iv{\FC \iv{\xi}, \ci{\xi_{a}}{\xi_{b}}}} \, .
\end{eqnarray}
Both $\CEN{\FC, \xi}$ and $\RMS{\FC, \xi}$ are undefined when $\MOM{\FC, \xi}{0} \iv{\FC \iv{\xi}, \ci{\xi_{a}}{\xi_{b}}} = 0$. 

One measure of the pulse duration, as evaluated over a time interval $\ci{t_{a}}{t_{b}}$ at a point $\ZR$, is 
\begin{equation}
\TU = 2 \RMS{\UE, t} \iv{\UE \iv{\ZR, t}, \ci{t_{a}}{t_{b}}} \, ,
\end{equation}
where the electromagnetic energy density
\begin{equation}
\UE \iv{z, t} = \Eo \av{\EF \iv{z, t}}^2 + \Mo \av{\HF \iv{z, t}}^2 \, .
\end{equation}
Alternatively, we could compute the RMS deviations from the centroid of $\Sz \iv{z, t}$ or the intensity $\av{\EF \iv{z, t}}^2$ with respect to time. Another measure of the pulse duration, called the equivalent duration~\cite{R.N.Bracewell-2000}, is $\MOM{\UE, t}{0}$ computed at a particular $\ZR$ over an interval $\ci{t_{a}}{t_{b}}$, and divided by the peak value of $\UE \iv{\ZR, t}$ over that interval, i.e.,
\begin{equation}
\TP = \nd{\MOM{\UE, t}{0} \iv{\UE \iv{\ZR, t}, \ci{t_{a}}{t_{b}}}}{\mv \iv{\UE \iv{\ZR, t}, \ci{t_{a}}{t_{b}}}} \, ,
\end{equation}
where $\mv$ indicates the maximum value of a function over the chosen time interval. The correlation duration $\TX$ is defined by~\cite{R.N.Bracewell-2000}
\begin{equation}
\TX = \nd{\int_{-\oo}^{\oo} \int_{-\oo}^{\oo} \UE \iv{\ZR, t} \UE \iv{\ZR, t + t'} dt' dt}{\MOM{\UE^{2}, t}{0} \iv{\UE^{2} \iv{\ZR, t}, \iv{-\oo, \oo}}} \, .
\end{equation}
We decided to use $\TP$, $\TU$, and $\TX$ as measures of pulse duration in this study. 

There are many measures of the pulse speed, including the group speed and the centrovelocity~\cite{R.L.Smith-1970}. We calculated three measures of the speed of the pulse through the film: the average speed of the pulse's peak value of $\UE \iv{\ZR, t}$, the average speed of $\CEN{\UE, t}$, and the average correlation velocity~\cite{S.C.Bloch-1977}. The energy density $\UE \iv{\ZR, t}$ of the transmitted pulse was recorded over the interval $\ci{t_{a}}{t_{b}}$ at some location $\ZR > \ZB{R}$. Then the time 
\begin{equation}
t_{p} = t_{\textrm{Max}} \iv{\UE \iv{\ZR, t}, \ci{t_{a}}{t_{b}}}
\end{equation}
where the function $t_{\textrm{Max}}$ indicates the time at which the maximum value of its argument occurs, the time 
\begin{equation}
t_{u} = \CEN{\UE, t} \iv{\UE \iv{\ZR, t}, \ci{t_{a}}{t_{b}}} \, ,
\end{equation}
and the time 
\begin{equation}
t_{c} = \TD + t_{\textrm{Max}} \iv{\int_{0}^{\oo} \UE \iv{0, t'} \UE \iv{\ZR, t + t'} dt'}
\end{equation}
were found. From these data, the average speeds $\SP$, $\SU$, and $\SX$ were calculated from the formula
\begin{equation}
\nd{c_{\ss}}{\Co} = \nd{\ZB{R} - \ZB{L}}{\Co \ld( t_{\ss} - \TD \rd) - \ld( z_{r} - \ld( \ZB{R} - \ZB{L} \rd) \rd)} \, , \quad \ss = p, u, c \, .
\end{equation}
Thus the speeds $\SP$, $\SU$, and $\SX$ are the thickness of the chiral STF divided by the computed times $t_{p}$, $t_{u}$, and $t_{c}$, less the time over which the pulse traverses the vacuous regions $0 < z < \ZB{L}$ and $\ZB{R} < z < \ZR$. We note that these measures are average speeds because they are in the form of a finite distance over a finite time interval, and hence do not reveal variations in the speed of the pulse as it enters, traverses, and exits the chiral STF. For the foregoing calculations, we approximated integrals with the rectangular rule~\cite{Y.Jaluria-1996}.

The measures of pulse duration and average speed have different physical meanings and concomitant advantages and disadvantages. The equivalent duration $\TP$ is the duration a rectangular pulse would have, if it possessed the same peak value of $\UE \iv{\ZR, t}$ and the same total energy per unit area as the pulse under consideration. Its advantage lies with its simplicity; however, it depends crucially on the correct determination of the peak value of $\UE \iv{\ZR, t}$. This same problem is a disadvantage of $\SP$, which is the average speed of the peak of the pulse through the chiral STF. This measure of the speed can even become multi--valued if the peak value of $\UE \iv{\ZR, t}$ occurs at more than one local maximum (like the twin humps of a Bactrian camel).

The duration $\TU$ is the RMS deviation of $\UE \iv{\ZR, t}$ from its first moment, normalized to the total energy per unit area of the pulse. It is analogous to the concept of radius of gyration in mechanics. We call $\SU$, the average speed of $\CEN{\UE, t}$, the center--of--energy speed. It is analogous to the  concept of the velocity of a body's center of mass in mechanics, and is closely related to the centrovelocity and a redefinition of the group velocity\footnote{The centrovelocity is the velocity of $\CEN{\av{\EF}^2, t} \iv{\av{\EF}^2 \iv{\ZR, t}, \ci{t_{a}}{t_{b}}}$~\cite{R.L.Smith-1970}.}, as discussed elsewhere~\cite{R.L.Smith-1970}. The advantage of these measures is that they depend on $\UE \iv{\ZR, t}$ over the whole pulse, and not just on the peak value of $\UE \iv{\ZR, t}$. However, it is possible for the center of energy to be located at a value of $z$ at which $\UE \iv{z, t} = 0$~\cite{S.C.Bloch-1977}. 

The correlation duration $\TX$ is a measure of how tightly $\UE \iv{\ZR, t}$ is concentrated in time. The correlation velocity $\SX$ is the average speed of the peak of the correlation of the incident pulse with the transmitted pulse. Like $\SP$, it depends heavily on the shape of the transmitted pulse. In some cases, it could become a multi--valued function of $\ZR$ and the chosen time interval $\ci{t_{a}}{t_{b}}$. However, it does provide a way to track the propagation of energy in the pulse in a different way than $\SU$, and it is fairly robust~\cite{S.C.Bloch-1977}. 

All six of the quantities we have defined in this subsection are amenable to experimental determination, for example by frequency--resolved optical gating (FROG)~\cite{R.Trebino-2000}. In our definitions of the measures, we have chosen to use $\UE \iv{\ZR, t}$ instead of $\av{\EF \iv{\ZR, t}}^{2}$, for example, because the former is a Lorentz--invariant quantity.

\section{Results and discussion} \label{S: Results and discussion}

We now turn to a discussion of the results of our calculations.

\subsection{Constitutive and excitation parameters} \label{S: Constitutive and excitation parameters}

The chiral STF we studied possesses the following constitutive parameters: $\OS{1} = 0.52$, $\OS{2} = 0.42$, $\OS{3} = 0.40$, $\AP{1,2,3} = 100$, $\RW{1} = 290$~nm, $\RW{2,3} = 280$~nm, $\HP{c} = 200$~nm, and $\AR = 20^{\circ}$. The film is structurally right--handed, i.e., $\SH = +1$, and we chose it to be either linear (i.e., $p_{nl} = 0$) or nonlinear ($p_{nl} = 3 \op 10^{-24}$~m$^{2}$/V$^{2}$), in accord with simple order--of--magnitude estimates~\cite{R.W.Boyd-1999}. We fixed $\ZB{L} = 30$~$\mu$m and $\ZB{R} = 34$~$\mu$m; thus, the chiral STF is ten pitches thick. The variable $z$ was discretized into intervals of $\dz = 2$~nm, and the variable $t$ into intervals of $\dt = 6.34 \op 10^{-3}$~fs. The center wavelength of the Bragg regime of the linear chiral STF is approximately~$516$~nm, and its full--width half--maximum bandwidth is approximately~$27$~nm~\cite{J.B.GeddesIII-2001A, A.Lakhtakia-1999}, though we note that both these quantities may change slightly when the chiral STF is nonlinear~\cite{J.B.GeddesIII-2003}. For all the results presented here, the incident pulse has the time constant $\TC = 2$~fs and time delay $\TD = 8$~fs. We calculated for three different combinations of material nonlinearity and incident pulse energy per unit area: 
\begin{center}
\begin{tabular}{ll}
Case L:  & a linear film, with $\UT = 1 \op 10^6$~J/m$^{2}$, \\
Case N1: & a nonlinear film, with $\UT = 1 \op 10^6$~J/m$^{2}$, or \\
Case N2: & a nonlinear film, with $\UT = 2 \op 10^{6}$~J/m$^{2}$.
\end{tabular}
\end{center}

Two indexes of refraction that govern the speed of monochromatic plane waves through a linear chiral STF are~\cite{A.Lakhtakia-2005} $\Nc \iv{\WL} = \sr{\Ec \iv{\WL}}$ and $\Nd \iv{\WL} = \sr{\Ed \iv{\WL}}$ where
\begin{equation}
\hat{\epsilon}_{\ell} \iv{\WL} 
= 1 + \hat{\chi}_{\ell} \iv{\WL} 
= 1 + \nd{p_{\ell} \FQ_{\ell} \ld[ \FQ_{\ell}^2 \ld[ 1 + \ld( 2 \pi N_{\ell} \rd)^{-2} \rd] - \ld( \nd{2 \pi \Co}{\WL} \rd)^2 + i \ld( \nd{2 \Co \FQ_{\ell}}{N_{\ell} \WL}\rd) \rd]}{\ld[ \FQ_{\ell}^2 \ld[ 1 + \ld( 2 \pi N_{\ell} \rd)^{-2} \rd] - \ld( \nd{2 \pi \Co}{\WL} \rd)^2 \rd]^{2} + \ld( \nd{2 \Co \FQ_{\ell}}{N_{\ell} \WL} \rd)^{2}} \, , \quad \ell = 1, 2, 3 \, ,
\end{equation}
and
\begin{equation}
\Ed \iv{\WL} = \nd{\hat{\epsilon}_1\iv{\WL} \hat{\epsilon}_3\iv{\WL}}{\hat{\epsilon_1}\iv{\WL} \sin^2 \AR + \hat{\epsilon_3}\iv{\WL} \cos^2 \AR} \, ,
\end{equation}
where $\hat{}$ indicates the temporal Fourier transform and $\WL$ is the free--space wavelength. Plots of the real and imaginary parts of $\Nc \iv{\WL}$ and $\Nd \iv{\WL}$ are shown in Figure~\ref{FigA}. The corresponding phase velocities
\begin{eqnarray}
\Vpc & = & \Co / \re \iv{\Nc} \, , \\
\Vpd & = & \Co / \re \iv{\Nd} \, ,
\end{eqnarray}
and group velocities
\begin{eqnarray}
\Vgc & = & \Co / \re \iv{\Nc + \AF \pd{\Nc}{\AF}} \, , \\
\Vgd & = & \Co / \re \iv{\Nd + \AF \pd{\Nd}{\AF}} \, ,  
\end{eqnarray}
are shown in Figure~\ref{FigB}; here, $\AF = 2 \pi \Co / \WL$ and $\pd{}{\AF} \equiv \partial / \partial \AF$. These frequency--domain plots help to explain our time--domain results.

\subsection{Reflected and transmitted pulses} \label{S: Reflected and transmitted pulses}

The spatiotemporal evolution of the incoming pulse was computed on the spatial domain $z \in \ci{0}{60}$~$\mu$m, of which the chiral STF occupies $z \in \ci{30}{34}$~$\mu$m. Measures of pulse duration and speed were computed at $\ZR = 36$~$\mu$m over the time interval $\ci{0}{190}$~fs. The wavelength $\CW$ of the carrier plane wave was set to either $415$, $515$, or $615$~nm; and the polarization state of the carrier plane wave was chosen as either LCP or RCP. At visible wavelengths, the selected linear constitutive properties yield refractive indexes that are commonplace in optics~\cite{H.A.Macleod-2001}.

Let us now examine the effects of the chiral STFs on pulses reflected from and transmitted through them. We are interested mainly in trends in the pulse shape, duration, and speed as the pulse energy per unit area, carrier wavelength and polarization state, and the nonlinearity of the chiral STF are varied, as opposed to particular values that the duration, etc., may take on as those parameters are varied. The shapes of the incident pulses, as exemplified by snapshots of their energy density in Figure~\ref{Fig1}, are independent both of the carrier wavelength and carrier polarization state. The incident pulses traversed vacuum ($z \in \ci{0}{30}$~$\mu$m), were scattered by the STF, and eventually became transmitted signals in the vacuous region $z \in \ci{34}{60}$~$\mu$m and reflected signals in the region $z \in \ci{0}{30}$~$\mu$m. 

Let us begin with Case~L. Both transmitted and reflected pulses are captured in the snapshots of $\UE \iv{z, t}$ at time $t = 190$~fs in Figure~\ref{Fig2}, where an important trend can be perceived. Irrespective of the polarization state of the carrier plane wave, the peak energy density of the transmitted pulse increases with increasing carrier wavelength. That is, the peak value of $\UE \iv{z, t}$, $z > \ZB{R}$, is greater when $\CW = 615$~nm than when $\CW = 515$ or $415$~nm for a given carrier polarization state. This trend applies in reverse to the lengths (i.e., $z$--extents) of the transmitted pulses: the transmitted pulses are longer when $\CW$ takes on lower values. 

The effects of structural handedness are most evident in the plots of the reflected pulses shown in Figure~\ref{Fig3}, which contains plots of $\UE \iv{z, t}$ as a function of time at $z = 27.6$~$\mu$m. The incident pulse is either LCP or RCP, and $\CW = 415$, $515$, or $615$~nm. The energy densities of the reflected pulses (which begin at $t \approx 110$~fs) are much smaller than those of the incident pulses.

The effects of the circular Bragg phenomenon are evident in Figure~\ref{Fig3}. In the bottom row of plots, for which the conditions for the circular Bragg phenomenon are satisfied~\cite{J.B.GeddesIII-2001A}, the primary reflected pulse possesses both a larger head and longer tail \emph{than} when the incident carrier plane wave is LCP (top row of plots). The longer tail is due to the bleeding of energy from the refracted pulse inside the film to the reflected pulse~\cite{J.B.GeddesIII-2001A}, and it is most pronounced when $\CW = 515$~nm, i.e. when the carrier wavelength lies in the Bragg regime. 

A secondary pulse appears in all six cases beginning at $t \approx 155$~fs. This smaller pulse is the result of the reflection from the chiral STF / vacuum interface at $z = \ZB{R}$. When the incident light is RCP, $\UE \iv{z, t}$ at $z < \ZB{L}$ oscillates rapidly in both the head of the primary pulse and the secondary pulse, a phenomenon that is due to interference between LCP and RCP components of the reflected light~\cite{J.B.GeddesIII-2003}. The tail of the primary reflected pulse does not exhibit such large oscillations because a superposition of LCP and RCP light of substantially the same energy densities is needed to create them~\cite{J.B.GeddesIII-2003}. The LCP component of the reflected pulse is of short duration, on the order of the duration of the incident pulse. The pulse--bleeding effect only operates on the fraction of incident light in the Bragg regime that is RCP, and so the tail of the primary reflected pulse contains mostly RCP light but little to no LCP light when the circular Bragg phenomenon occurs.

\begin{figure}[f]
\begin{center}
\includegraphics[width = 2 in]{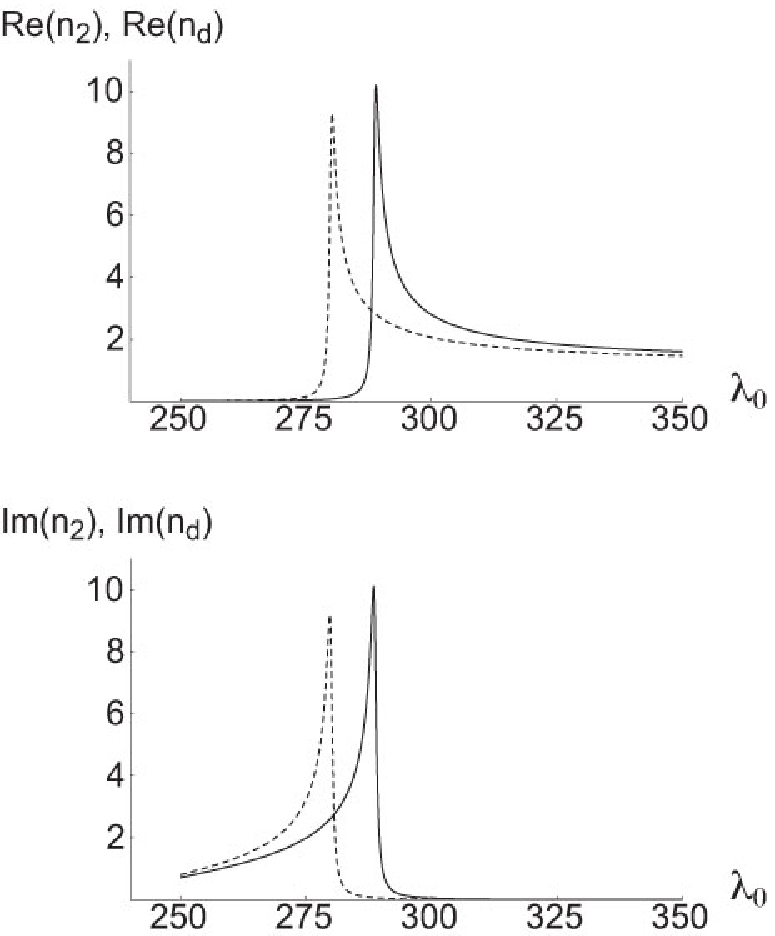}
\end{center}
\caption{Plots of $\re \iv{\Nc}$ (top, dashed), $\re \iv{\Nd}$ (top, solid), $\im \iv{\Nc}$ (bottom, dashed), and $\im \iv{\Nd}$ (bottom, solid) as functions of free--space wavelength $\WL$ (in~nm). Note that these quantities decrease with increasing $\WL$ for $\WL > 300$~nm.} \label{FigA}
\end{figure}

Corresponding plots for the nonlinear Cases N1 and N2 are shown in Figures~\ref{Fig4} and~\ref{Fig5}, respectively. While still small compared to the incident pulses', the peak values of $\UE \iv{z, t}$ in the heads of the primary reflected pulses increase from Figures~\ref{Fig3} to~\ref{Fig5} for given carrier wavelength and polarization state. This is due to the increasing effects of nonlinearity in Case~N1 as compared to Case~L, and in Case~N2 as compared to Case~N1. An understanding emerges from the frequency domain~\cite{J.B.GeddesIII-2001A, J.Wang-2002, J.B.GeddesIII-2003} as follows: for the type of nonlinearity we studied, a greater value of $\UT$ creates a greater refractive--index mismatch between vacuum and the chiral STF because of the multiplicative character of the nonlinearity in~(\ref{E: Causal response})~\cite{J.B.GeddesIII-2003}. The refractive index of vacuum is unity, and we can see from Figure~\ref{FigA} that the real parts of $\Nc \neq 1$ and $\Nd \neq 1$ are decreasing functions of carrier wavelength at carrier wavelengths greater than 300~nm. When the chiral STF is nonlinear, the differences between the refractive indexes of the chiral STF and unity shall become even greater, and those differences will be greatest when the intensity of the beam is highest. Hence, the reflection at the pulse head is greater than otherwise. 

The peak values of $\UE \iv{z, t}$ in the secondary pulses are not appreciably different from Case~L to Case~N1 to Case~N2 (at least when normalized by the value of $\UT$, as was done for all the figures). Evidently, absorption by the chiral STF reduces the peak energy density of the primary refracted pulse in Case~N2 to a greater degree than in Case~N1 and in Case~N1 to a greater degree than in Case~L~\cite{J.B.GeddesIII-2003}. In the frequency domain, absorption manifests itself through positive values of $\im \iv{\Nc}$ and $\im \iv{\Nd}$, as seen in Figure~\ref{FigA}. By the time the primary refracted pulses reach $\ZB{R}$, their energy densities drop such that the refractive--index mismatch between vacuum and chiral STF across $z = \ZB{R}$ is reduced between Cases~L, N1, and N2. Differences in energy density that remained were further reduced on the return trip of the secondary pulse from $z = \ZB{R}$ to $z = \ZB{L}$. 

Figures~\ref{Fig6} to~\ref{Fig8} contain plots of $\UE \iv{z, t}$ as functions of time recorded at $z = 36$~$\mu$m; these are the transmitted pulses for Cases L, N1, and N2, respectively. The peak values of $\UE \iv{z, t}$ increase, for given carrier polarization state, as $\CW$ increases, an effect that is attributable to the presence of the absorption resonances at ultraviolet wavelengths. The frequency domain behavior of the absorption resonances is shown in plots of $\im \iv{\Nc}$ and $\im \iv{\Nd}$ in Figure~\ref{FigA}. Other parameters being equal, more of the bandwidth of a pulse with lower $\CW$ overlaps the bandwidth of the absorption resonances of the chiral STF than the bandwidth of a pulse with higher $\CW$, and hence more absorption occurs in the former case. For given $\CW$ and carrier polarization state, the peak values of $\UE \iv{z, t}$---again normalized to $\UT$---drop from Figures~\ref{Fig6} to~\ref{Fig8}. Notice that when the incident carrier plane wave is RCP, but not LCP, the transmitted pulses tend to have a lobed or humped structure, and that the number and sizes of the lobes increase from Case L to Case N1 to Case N2. Thus, the presence of the chosen nonlinearity distorts the shapes of the transmitted pulses and tends to make them highly asymmetric, but the effect is prominent only when the structural handedness of the film matches the sense of the circular polarization state of the incident carrier plane wave.

\subsection{Durations of transmitted pulses}

What effects does a change in carrier wavelength have on the durations of pulses transmitted through a chiral STF? Quantification of the effects of the carrier wavelength, carrier polarization state, and the nonlinearity of the chiral STF on $\TP$, $\TU$, and $\TX$ is provided through Figure~\ref{Fig9}, wherein plots of those quantities for thirteen values of $\CW$ ranging from~395 to~635~nm are shown. The incident carrier plane wave is again either LCP or RCP, and Cases L, N1, and N2 are again considered. The overall trend in the plots is a decrease in pulse duration with increasing carrier wavelength. As the carrier wavelength decreases, the material exhibits greater dispersion due to proximity to the absorption resonances. The transmitted pulse is consequently of longer duration, because light of longer free--space wavelength has both greater phase and group velocities in the chiral STF than light of shorter $\WL$, as can be seen in Figure~\ref{FigB}. This is because $\Nc \iv{\WL}$ and $\Nd \iv{\WL}$ drop with increasing $\WL$ above the resonance wavelengths $\lambda_{1,2,3}$, as can be seen in Figure~\ref{FigA}. Therefore the chosen pulses, which do not possess any negative chirp in their carrier wavelengths, tend to spread out as they propagate through the chiral STF. 

Although that is the overall trend, there is an exception: there appears to be a local maximum for $\TU$ when $\CW \approx 515$~nm and the incident carrier plane wave is RCP. We believe that this effect is due to the circular Bragg phenomenon, because the local maximum appears in the vicinity of the center wavelength of the Bragg regime, and it does not appear when the incident carrier plane wave is LCP. Additionally, for the same $\CW$ and carrier polarization state, the duration of the pulses tends to increase from Case~L to Case~N1 to Case~N2. 

\subsection{Average transmission speed through chiral STF}

\begin{figure}[f]
\begin{center}
\includegraphics[width = 2 in]{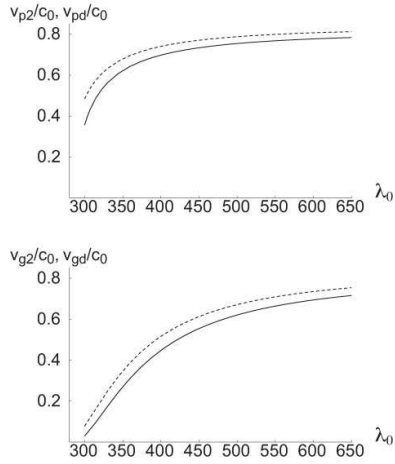}
\end{center}
\caption{Plots of $\Vpc / \Co$ (top, dashed), $\Vpd / \Co$ (top, solid), $\Vgc / \Co$ (bottom, dashed), and $\Vgd / \Co$ (bottom, solid) as function of free--space wavelength $\WL$ (in~nm). Note that these quantities increase with increasing $\WL$.} \label{FigB}
\end{figure}

\begin{figure}[f]
\begin{center}
\includegraphics[width = 4 in]{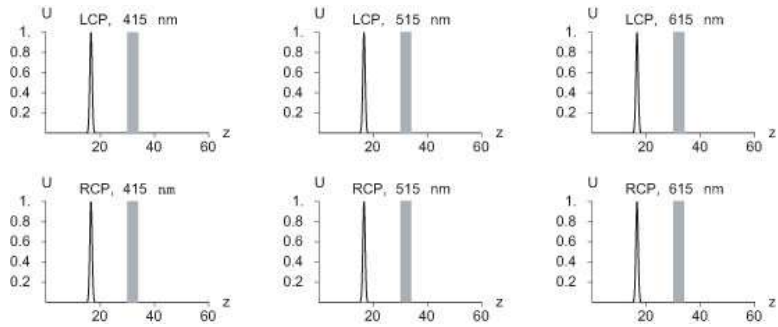}
\end{center}
\caption{Plots of $\UE \iv{z, t}$ (in arbitrary units) of the incident pulses as a function of $z$ (in~$\mu$m) for $\CW = 415$~nm (left), $515$~nm (center), and $615$~nm (right) recorded at time $t = 63.4$~fs. The incident carrier plane wave is either LCP (top) or RCP (bottom); note that the plots are identical regardless of carrier wavelength or polarization. The chiral STF, indicated by the shaded area, occupies the region $z \in \ci{30}{34}$~$\mu$m. The values of $\UE \iv{z, t}$ have been multiplied by a factor of $\Co \TC \sr{\pi} / \UT$; hence, the shapes of the incident pulses in these plots are independent of $\UT$ also.} \label{Fig1}
\end{figure}

\begin{figure}[f]
\begin{center}
\includegraphics[width = 4 in]{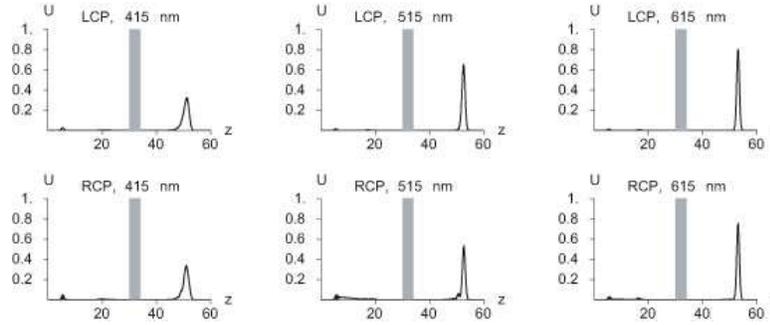}
\end{center}
\caption{Same as Figure~\ref{Fig1}, except that the data are recorded at time $t = 190$~fs, and so reflected (left of film) and transmitted (right of film) pulses are captured. The chiral STF is structurally right--handed and linear (Case L).} \label{Fig2}
\end{figure}

\begin{figure}[f]
\begin{center}
\includegraphics[width = 4 in]{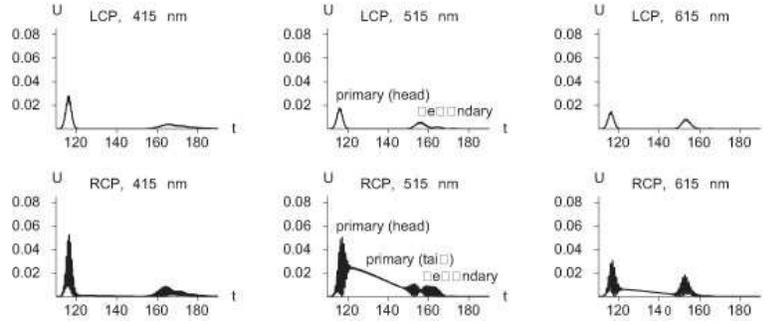}
\end{center}
\caption{Plots of $\UE \iv{z, t}$ (in arbitrary units) as a function of time t (in~fs) for $\CW = 415$~nm (left), $515$~nm (center), and $615$~nm (right) recorded at $z = 27.6$~$\mu$m. The incident carrier plane wave is either LCP (top) or RCP (bottom). The primary and secondary reflected pulses are captured in the record. The chiral STF is structurally right--handed and linear (Case L). The values of $\UE \iv{z, t}$ are multiplied by a factor of $\Co \TC \sr{\pi} / \UT$, where $\UT = 1 \op 10^6$~J/m$^2$.} \label{Fig3}
\end{figure}

\begin{figure}[f]
\begin{center}
\includegraphics[width = 4 in]{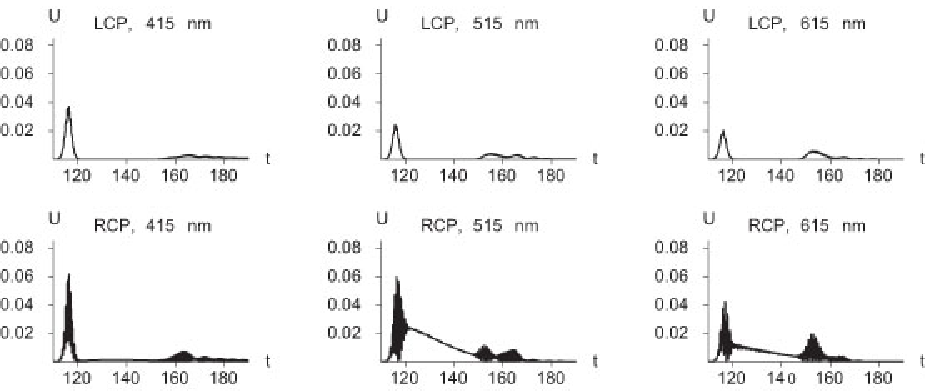}
\end{center}
\caption{Same as Figure~\ref{Fig3}, except that the chiral STF is nonlinear (Case N1).} \label{Fig4}
\end{figure}

\begin{figure}[f]
\begin{center}
\includegraphics[width = 4 in]{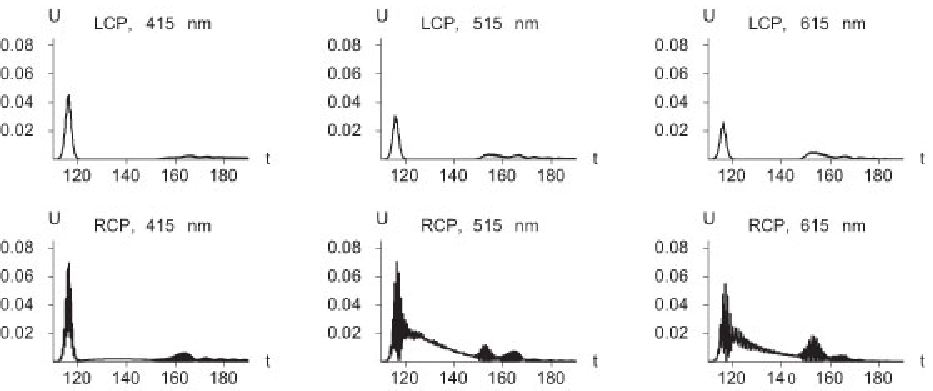}
\end{center}
\caption{Same as Figure~\ref{Fig4}, except that $\UT = 2 \op 10^{6}$~J/m$^2$ (Case N2).} \label{Fig5}
\end{figure}

\begin{figure}[f]
\begin{center}
\includegraphics[width = 4 in]{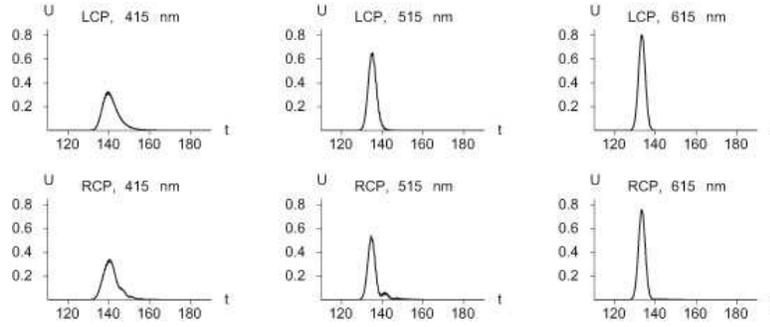}
\end{center}
\caption{Plots of $\UE \iv{z, t}$ (in arbitrary units) as a function of time t (in~fs) for $\CW = 415$~nm (left), $515$~nm (center), and $615$~nm (right) recorded at $z = 36$~$\mu$m, and so the transmitted pulses are captured in the record. The incident carrier plane wave is either LCP (top) or RCP (bottom), and the chiral STF is structurally right--handed and linear (Case L). The values of $\UE \iv{z, t}$ are multiplied by a factor of $\Co \TC \sr{\pi} / \UT$, where $\UT = 1 \op 10^6$~J/m$^{2}$.} \label{Fig6}
\end{figure}

\begin{figure}[f]
\begin{center}
\includegraphics[width = 4 in]{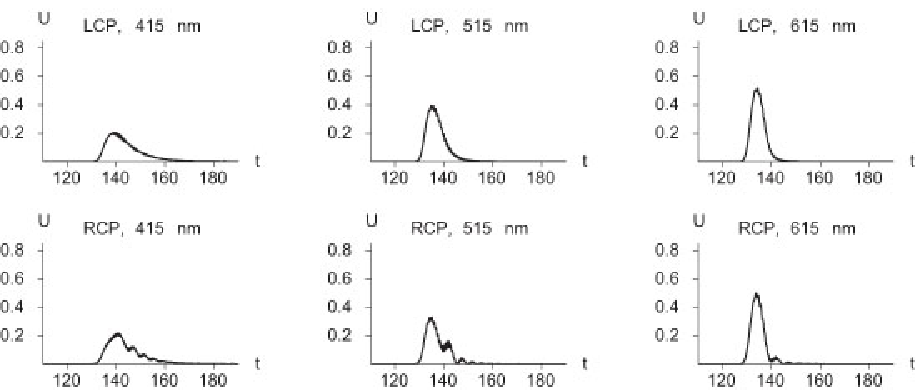}
\end{center}
\caption{Same as Figure~\ref{Fig6}, except that the chiral STF is nonlinear (Case N1).} \label{Fig7}
\end{figure}

\begin{figure}[f]
\begin{center}
\includegraphics[width = 4 in]{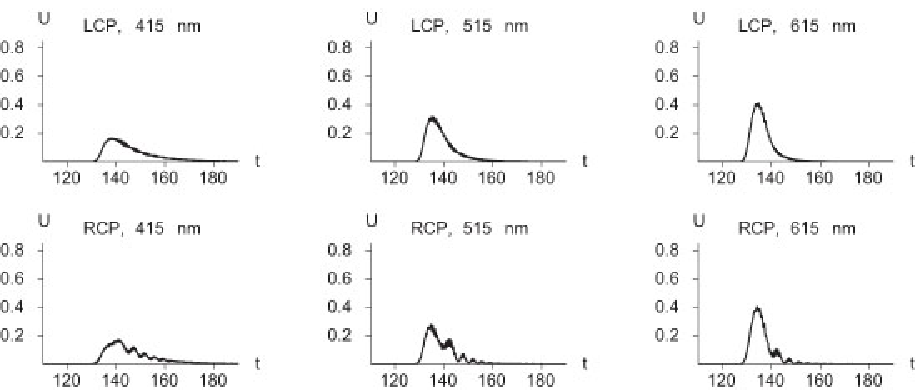}
\end{center}
\caption{Same as Figure~\ref{Fig7}, except that $\UT = 2 \op 10^{6}$~J/m$^2$ (Case N2).} \label{Fig8}
\end{figure}

\begin{figure}[f]
\begin{center}
\includegraphics[width = 4 in]{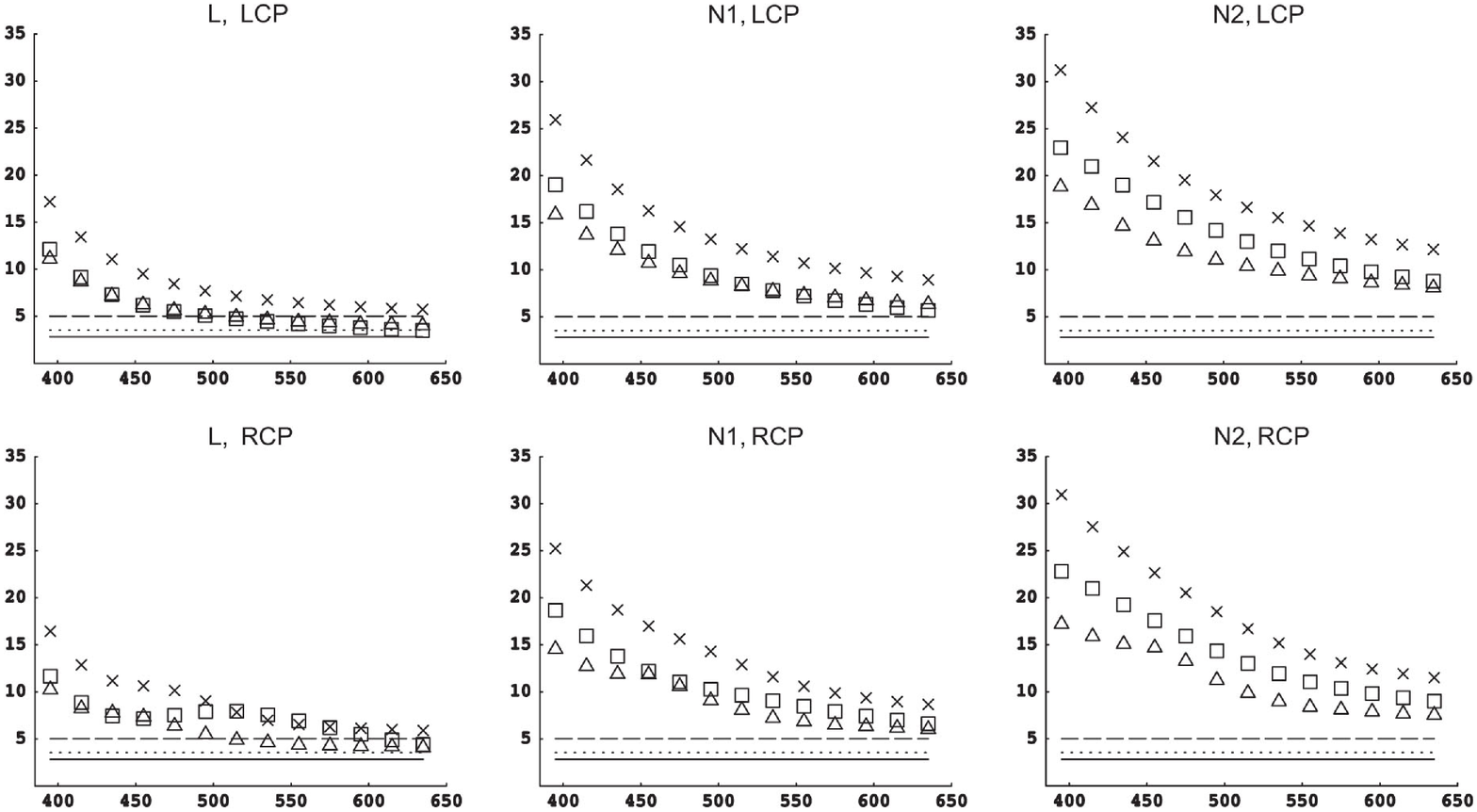}
\end{center}
\caption{Durations $\TP$ ($\triangle$, in~fs), $\TU$ ($\Box$, in~fs), $\TX$ ($\ml$, in~fs) as functions of carrier wavelength $\CW$ (in~nm) for six different cases, evaluated at $\ZR = 36$~$\mu$m over the interval $\ci{0}{190}$~fs. The incident carrier plane wave is either LCP (top) or RCP (bottom). In addition, the chiral STF is either linear ($\NP = 0$) while $\UT = 1 \op 10^6$~J/m$^{2}$ (Case L, at left); nonlinear ($\NP = 3 \op 10^{-24}$~m$^2$/V$^{2}$) while $\UT = 1 \op 10^6$~J/m$^{2}$ (Case N1, in center); or nonlinear ($\NP = 3 \op 10^{-24}$~m$^2$/V$^{2}$) while $\UT = 2 \op 10^6$~J/m$^{2}$ (Case N2, at right). The lines at the bottom of each plot indicate the value of $\TP$ (dotted), $\TU$ (solid), and $\TX$ (dashed) for the incident pulses.} \label{Fig9}
\end{figure}

\begin{figure}[f]
\begin{center}
\includegraphics[width = 4 in]{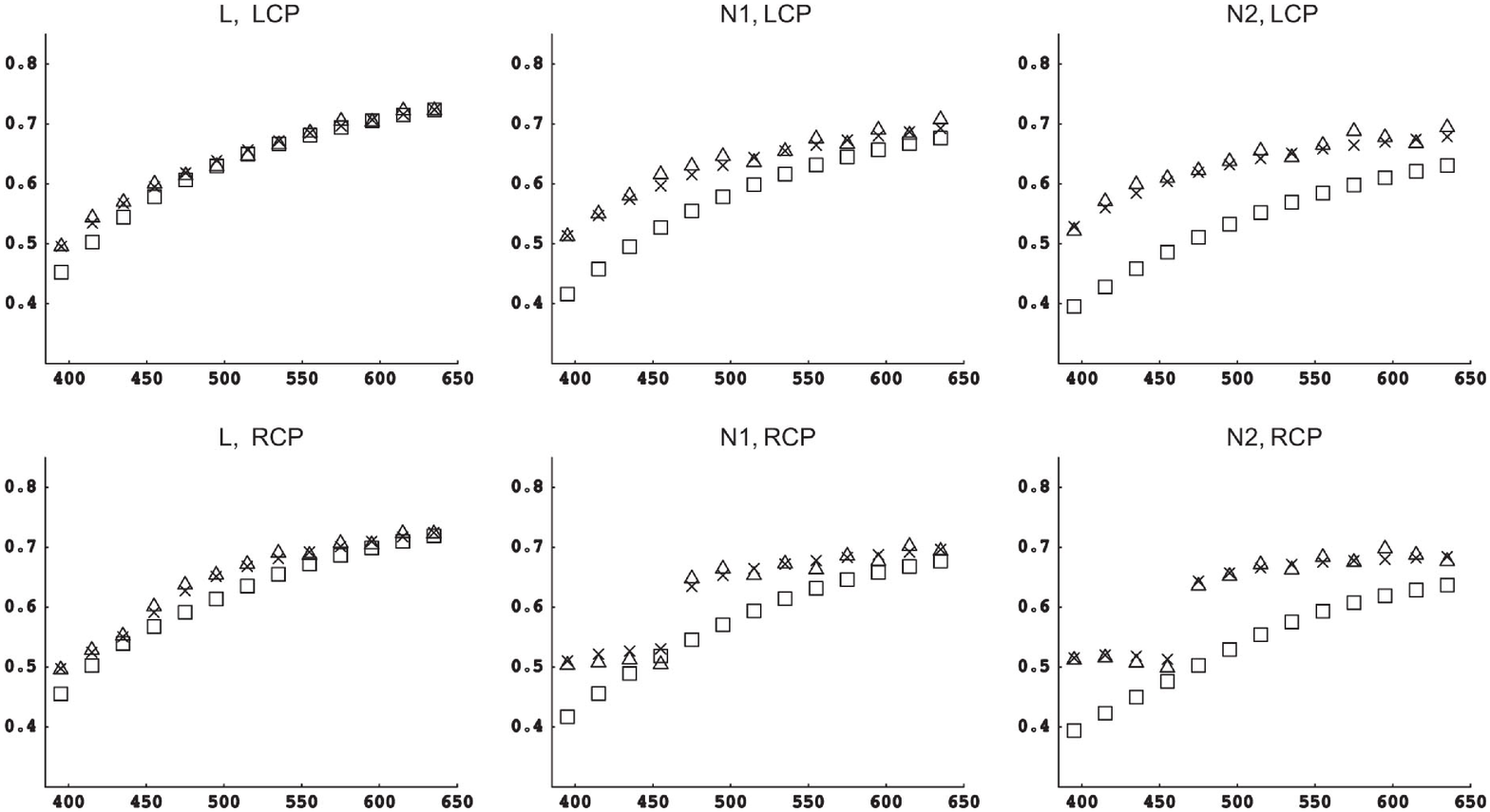}
\end{center}
\caption{Normalized average speeds $\SP/\Co$ ($\triangle$), $\SU/\Co$ ($\Box$), and $\SX/\Co$ ($\ml$) as functions of carrier wavelength $\CW$ (in~nm) for six different cases, evaluated at $\ZR = 36$~$\mu$m over the interval $\ci{0}{190}$~fs. The incident carrier plane wave is either LCP (top) or RCP (bottom). In addition, the chiral STF is either linear ($\NP = 0$) while $\UT = 1 \op 10^6$~J/m$^{2}$ (Case L, at left); nonlinear ($\NP = 3 \op 10^{-24}$~m$^2$/V$^{2}$) while $\UT = 1 \op 10^6$~J/m$^{2}$ (Case N1, in center); or nonlinear ($\NP = 3 \op 10^{-24}$~m$^2$/V$^{2}$) while $\UT = 2 \op 10^6$~J/m$^{2}$ (Case N2, at right).} \label{Fig10}
\end{figure}

Plots of $\SP$, $\SU$, and $\SX$ in Figure~\ref{Fig10} show that they all tend to increase  with increasing carrier wavelength, but decrease slightly with increasing nonlinearity. These two trends are due to the presence of the absorption resonances in the ultraviolet regime, and the tendency of the nonlinearities to enhance those resonances. As is clear from Figure~\ref{FigA}, the principal refractive indexes of the chiral STF are higher for wavelengths closer to the resonances, and hence pulsed plane waves with lower $\CW$ are slowed more by the film than pulsed plane waves with higher $\CW$, as can be seen in Figure~\ref{FigB}.

The presence of nonlinearity increases the refractive indexes because of its multiplicative nature in~(\ref{E: Causal response}), and hence pulsed plane waves with higher $\UT$ are slowed more than pulsed plane waves with lower $\UT$ when propagating through a chiral STF. Both $\SP$ and $\SX$ are heavily affected by pulse shaping by the chiral STF which is manifested as the alteration of the peaks of the transmitted pulses. For example, notice the jumps in these quantities that occur between $\CW = 455$~nm and $\CW = 475$~nm for Cases~N1 and~N2 when the incident carrier plane wave is RCP. The jumps occur due to the appearance of the lobed structure of the transmitted pulses mentioned in Section~\ref{S: Reflected and transmitted pulses}; and, in the case of $\SX$, also due to the the short duration of the incident pulse as compared to the transmitted one. The calculations of $\SP$ and $\SX$ require finding the peak value of either $\UE \iv{\ZR, t}$ or of its correlation with the $\UE \iv{0, t}$; and so, if one lobe suddenly becomes larger than another, the peak of the transmitted pulse has to shift. This shift creates a jump in both $\SP$ and $\SX$. Pulse shaping is to be expected to be particularly prominent when pulses of high peak powers pass through nonlinear chiral STFs, because the effects of nonlinearity are going to be very pronounced then.

\section{Concluding remarks} \label{S: Concluding remarks}

As a result of our calculations, we offer the following conclusions that may be useful to future designers of optical pulse shapers fabricated with chiral STFs.

\begin{itemize}
\item Increased nonlinearity reshapes pulses more---as is evident in the increased formation of a lobed structure---when the carrier polarization state of the incident pulse has the same sense as the structural handedness of the chiral STF. Attenuation of the transmitted pulse increases when the carrier wavelength lies closer to the absorption resonances, which is also confirmed by frequency--domain results~\cite{J.Wang-2002, A.Lakhtakia-1999}.
\item For specified carrier polarization state, $\UT$, and $\NP$, the durations $\TP$, $\TU$, and $\TX$ all tend to increase with decreasing $\CW$---except for $\TU$ in the case of a linear chiral STF under conditions in which the circular Bragg phenomenon is exhibited. Therefore a chiral--STF--based device could change the durations of pulses, contingent on the polarization state of the incident carrier plane wave.
\item For specified $\CW$ and polarization state of the incident carrier plane wave, the transmitted pulse duration tends to increase with increasing nonlinear effects. Therefore, any pulse--shaping device based on nonlinear chiral STFs can only be expected to work over a certain range of incident pulse intensities.
\item For a given carrier polarization state, $\UT$, and $\NP$, the speed $\SU$ tends to increase with increasing $\CW$. Therefore chiral STFs can function as wavelength--selective delay elements for ultrashort optical pulses.
\item For specified $\CW$ and polarization state of the incident carrier plane wave, the speeds $\SP$, $\SU$, and $\SX$ tend to decrease with increasing nonlinear effects.
\end{itemize}

To conclude, let us note that our presented technique can accommodate other types of nonlinearities---specifically, nonlinearities that are not simply multiplicative but must be incorporated through multiple convolutions, which appear necessary for modelling the nonlinear response of dielectrics to attosecond pulses~\cite{R.W.Hellwarth-1977}. However, the computational capabilities then required will be significantly greater than the ones commonly available today.

\bigskip

{\bf Acknowledgments:} We thank the Pittsburgh Supercomputing Center for computer resources used to develop the computer programs used in this work. Joseph B. Geddes III thanks both the National Science Foundation (NSF) for an NSF Graduate Research Fellowship and SPIE for a SPIE Educational Scholarship. We are grateful for the reviews of two anonymous referees and an editor whose comments helped us improve the presentation of our work.



\newpage

\end{document}